\documentclass{revtex4}

\usepackage{graphicx}
\usepackage{amsmath}
\usepackage{amssymb}
\usepackage{color}
\usepackage{epstopdf}

\begin{document}
\title{Dynamical Effects of Multiplicative Feedback on a Noisy System}

\author{Giuseppe Pesce}
\affiliation{Dipartimento di Scienze Fisiche, Universit\'{a} di Napoli `Federico II', Complesso Universitario Monte S Angelo Via Cintia, I-80126 Napoli, Italy}
\author{Austin McDaniel}
\author{Scott Hottovy}
\author{Jan Wehr}
\affiliation{Department of Mathematics and Program in Applied Mathematics, University of
Arizona, Tucson, Arizona 85721 USA}
\author{Giovanni Volpe}
\affiliation{Department of Physics, Bilkent University, Cankaya, Ankara 06800, Turkey}


\begin{abstract}
Intrinsically noisy mechanisms drive most physical, biological and economic phenomena, from stock pricing to phenotypic variability. Frequently, the system's state influences the driving noise intensity, as, for example, the actual value of a commodity may alter its volatility or the concentration of gene products may regulate their expression. All these phenomena are often modeled using stochastic differential equations (SDEs). However, an SDE is not sufficient to fully describe a noisy system with a multiplicative feedback, because it can be interpreted according to various conventions -- in particular, It\^{o} calculus and Stratonovitch calculus --, each of which leads to a qualitatively different solution. Which convention to adopt must be determined case by case on the basis of the available experimental data; for example, the SDE describing electrical circuits driven by a noise are known to obey Stratonovich calculus. Once such an SDE-convention pair is determined, it c
 an be employed to predict the system's behavior under new conditions. Here, we experimentally demonstrate that the convention for a given physical system may actually vary under varying operational conditions. We show that, under certain conditions, a noisy electric circuit shifts to obey It\^o calculus, which may dramatically alter the system's long term stability. We track such Stratonovich-to-It\^o transition to the underlying dynamics of the system and, in particular, to the ratio between the driving noise correlation time and the feedback delay time. We briefly discuss ramifications of our conclusions for biology and economics: the possibility of similar transitions and their dramatic consequences should be recognized and accounted for where SDEs are employed to predict the evolution of complex phenomena.
\end{abstract}

\maketitle

\section{Introduction}\label{sec:intro}

\begin{figure}
\includegraphics[width=10cm]{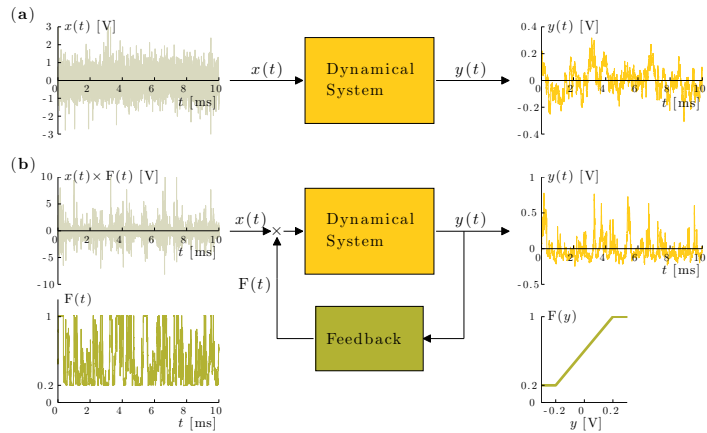}
\caption{{\bf Stochastic dynamical system without and with feedback.} (a) Schematic of a stochastic dynamical system: the system's status $y(t)$ evolves as the system is driven by a noisy input $x(t)$. (b) Same system with feedback $F(y)$: $x(t)$ is now modulated by $F(t)$ and $y(t)$ is clearly affected.
\label{fig1}}
\end{figure}

Mathematical models are often employed to predict the behavior and evolution of complex physical, chemical, biological and economic phenomena. 
Often more realistic mathematical models can be obtained by allowing for some randomness \cite{Oksendal}. In a dynamical system, for example, this randomness can be introduced by adding a noisy driving term, where a noise $x_t$ drives the evolution of the state $y_t$ of the dynamical system [Fig. \ref{fig1}(a)]. Similar models have been employed to describe a wide range of phenomena, from thermal fluctuations of microscopic objects \cite{Nelson} and the evolution of stock prices \cite{Bachelier1900} to heterogeneous response of biological systems to stimuli \cite{Blake2003} and stochasticity in gene expression \cite{Kaern2005}.

Intrinsically noisy phenomena are often modeled using stochastic differential equations (SDEs) \cite{Oksendal}. An SDE is obtained by adding some randomness to a deterministic dynamical system described by an ordinary differential equation (ODE) \cite{Strogatz}. A typical SDE has the form:
\begin{equation}\label{eq:SDE}
d y_t = G(y_t) \, dt + \sigma \, dW_t,
\end{equation}
where $G(y)$ is a function representing the deterministic response of the system, $W_t$ is a Wiener process representing the stochastic driving, and $\sigma$ is a scaling constant representing the intensity of the noise. Clearly, the term $\sigma\,dW_t$ is a mathematical model of the physical noise $x_t\,dt$. In particular, any real process has always a correlation time $\tau>0$, while $dW_t$ is strictly uncorrelated, i.e. $\tau=0$; therefore, the smaller the $\tau$ of a real process, the better it is approximated by $dW_t$ \cite{Kloeden}. We remark that Eq. (\ref{eq:SDE}) has a unique solution with a given initial condition $y_0$.  This solution satisfies the integral equation $y_t = y_0 + \int_0^tG(y_s)ds + \sigma W_t$ \cite{Oksendal}.

In many real phenomena, the system's state further influences the driving noise intensity [Fig. \ref{fig1}(b)]; for example, the volatility of a stock price may be altered by its actual value \cite{Hamao1990} or gene expression may be regulated by the concentration of its products \cite{Kaern2005}. This multiplicative feedback $F(y)$ leads us to consider an SDE with multiplicative noise:
\begin{equation}\label{eq:SDEmult}
d y_t = G(y_t)\,dt + \sigma\,F(y_t)\,dW_t.
\end{equation}
Unlike Eq. (\ref{eq:SDE}), the integration of Eq. (\ref{eq:SDEmult}) presents some difficulties because $W_t$ is a function of unbounded variation \cite{Oksendal}. The {\it stochastic integral} $\int_{0}^{T} f(y_t) \circ_\alpha dW_t \equiv  \lim_{n \to \infty} \sum_{n=0}^{N} f(y_{t_n})\Delta W_{t_n}$, where $t_n = \frac{n+\alpha}{N} T$ and $\alpha \in [0,1]$, leads to different values for each choice of $\alpha$ \cite{Karatzas, Sussmann1978}. Common choices are: the {\it It\={o} integral} with  $\alpha=0$ \cite{Ito1944}; the {\it Stratonovitch integral} with $\alpha=0.5$ \cite{Stratonovich1966}; and the {\it anti-It\={o}} or {\it isothermal integral} with $\alpha=1$ \cite{Klimontovich1990}. Alternative values of $\alpha$ may entail dramatic consequences; for example, a Malthusian population growth model with a noisy growth rate can lead either to extinction, if solved with $\alpha = 0$, or to exponential growth, if solved with $\alpha = 0.5$ \cite[Section 5.1]{Oksendal}. Ther
 efore, a complete model is 
 defined by the SDEs \emph{and} the relative convention, which must be determined on the basis of the available experimental data \cite{vanKampen1981}. Various preferences regarding the appropriate choice of $\alpha$ have emerged in various fields in which SDEs have been fruitfully applied. For example, $\alpha = 0$ is typically employed in economics \cite{Oksendal} and biology \cite{Turelli1977}, because of its property of ``not looking into the future,'' referring to the fact that, when the integral is approximated by a sum, the first point of each interval is used. $\alpha = 0.5$ naturally emerges in real systems with non-white noise, i.e. $\tau > 0$, e.g. the SDEs describing electrical circuits driven by a multiplicative noise \cite{Smythe1983}, as a consequence of the Wong-Zakai theorem, which states that, if the Wiener process is substituted by smooth process with $\tau \rightarrow 0$, the resulting SDE obeys the Stratonovich calculus \cite{Wong1969}. Finally, $\alpha =
  1$ naturally emerges in physical systems in equilibrium with a heat bath \cite{Ermak1978,Lancon2001,Volpe2010}. Other values of $\alpha$ have also been theoretically proposed \cite{Kupferman2004, Freidlin2004,Hottovy2012}. Clearly, from the modelling perspective the choice of the appropriate SDEs-convention pair is of critical importance, especially when the model is subsequently employed to predict the system's behavior under new conditions.

In this article, we experimentally demonstrate that the convention for a given physical system can actually vary under changing operational conditions. We show that the equation describing the behavior of an electric circuit with multiplicative noise, which usually obeys the Stratonovitch convention ($\alpha = 0.5$) \cite{Smythe1983}, crosses over to obey the It\^o convention ($\alpha = 0$), as certain parameters of the dynamical systems are changed. This transition is continuous, going through all intermediate values of $\alpha$ and we relate it by an explicit formula to the ratio between $\tau$ and the feedback delay time $\delta$, which is always present in any real system. Similar transitions have the potential of dramatically altering a system's long term behavior and, therefore, we argue their possibility should be taken into account in the modelling of systems with SDEs, which are widely employed in economics, biology and physics.

\section{System without feedback}

\begin{figure}
\includegraphics[width=10cm]{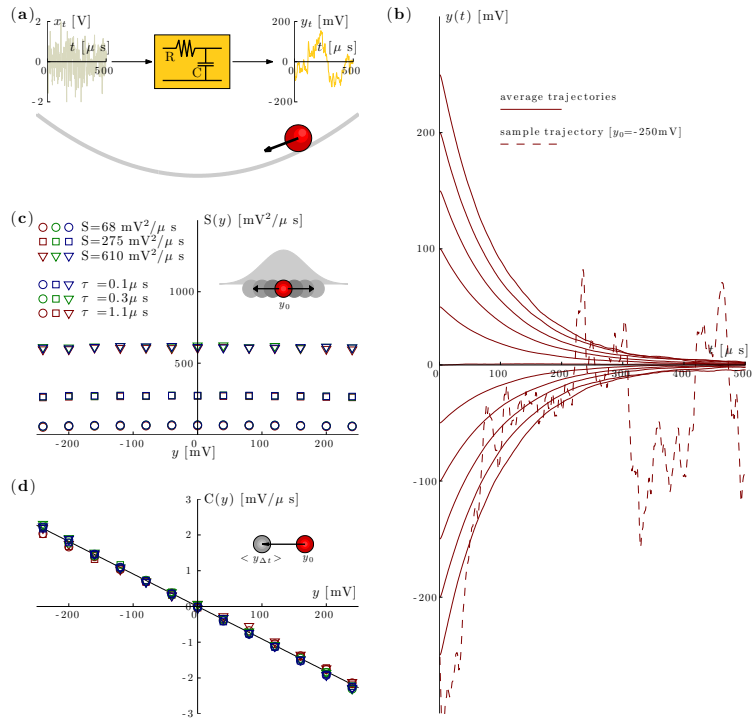}
\caption{{\bf Stochastic dynamical system without feedback.} (a) In our experiments, we employ a RC electric circuit driven by a noise $x_t$; the system's status $y_t$ effectively experiences a harmonic restoring $G(y) = -ky$ with $k = 1/RC$. (b) Sample trajectory of $y_t$ ($\tau = 1.1\,\mathrm{\mu s}$) with initial condition $y_0 = \mathrm{-250\, mV}$ (dashed line) and average of 1000 trajectories for various initial conditions (solid lines). (c) Diffusion $S(y)$ and (d) drift $D(y)$ of the system status for various intensities and correlation times ($\tau$) of the input noise. $S(y)$ is proportional to the variance of the system state change [inset in (c)] and $D(y)$ to its average [inset in (d)]. The solid line in (c) represents the harmonic restoring $G(y)$.
\label{fig2}}
\end{figure}

A  system near an equilibrium is often described as a harmonic oscillator, where a {\it restoring force}  brings the system back towards the equilibrium. Such harmonic oscillators are widely employed to describe the behavior of systems near their equilibria, from the swinging of pendula to the vibrations of atoms in crystals. In this work, as a paradigmatic experimental realization of an overdamped harmonic oscillator, we consider an RC electric circuit with resistance $R = 1\,\mathrm{k\Omega}$ and capacitance $C = 100\,\mathrm{nF}$; $x_t$ is the driving voltage (applied on the series RC) and $y_t$ the output voltage (measured on C) [Fig. \ref{fig2}(a)]. In order to approximate a Wiener process, we will always use a driving noise with a correlation time much shorter than the typical relaxation time of the circuit, i.e. $\tau \ll RC = 100\,\mathrm{\mu s}$. A detailed description of the circuit is given in the methods section. The output
  voltage $y$ experiences an elastic restoring force with elastic constant $k = 1/RC$ towards the $y = 0$ equilibrium state.

In order to understand qualitatively the behavior of our system, in Fig. \ref{fig2}(b) we consider the evolution of $y_t$ for a given initial condition $y_0$ and $\tau = 1.1\,\mathrm{\mu s}$. The dashed line illustrates a sample trajectory for $y_0 = -250\,\mathrm{mV}$: at the beginning $y_t$ decays towards the equilibrium $y = 0\,\mathrm{mV}$ and, afterwards, oscillates around the equilibrium, clearly demonstrating its stochastic nature. Averaging several such trajectories, we obtain solid lines corresponding to different $y_0$, which clearly show that the average trajectory moves towards the equilibrium regardless of $y_0$. 

The relevant SDE is Eq. (1) with $G(y) = -k y$ and $\sigma$ proportional to the intensity of the noise, i.e.:
\begin{equation}\label{eq:SDEcircuit}
dy_t = -k\,y_t\,dt + \sigma\,dW_t,
\end{equation}
where we remark that, since $\sigma$ is constant, the choice of $\alpha$ does not affect the solution and, therefore, the convention can be left undetermined.

In a very general sense, a system described by an SDE can be characterized by its stochastic diffusion $S(y)$ and its drift $D(y)$ \cite{vanKampen1981}. Letting the system evolve from an initial state $y$ for an infinitesimal time-step, $S(y)$ is proportional to the variance of the system's state change [inset in Fig. \ref{fig2}(c)] and $D(y)$ to its average [inset in Fig. \ref{fig2}(d)]. $S(y)$ and $D(y)$ can be obtained from an experimental discrete time-series $\{y_0, ... , y_{N-1}\}$ sampling the output signal at intervals $\Delta t$ as
\begin{equation}\label{eq:D2}
S(y) = \frac{1}{2 \Delta t} \left< (y_{n+1} - y_n)^2 \mid y_n \cong y \right>
\end{equation}
and
\begin{equation}\label{eq:D1}
D(y) = \frac{1}{\Delta t} \left< y_{n+1} - y_n \mid y_n \cong y \right>.
\end{equation}
Eqs. (\ref{eq:D2}) and (\ref{eq:D1}) are strictly true in the limit $\Delta t \rightarrow 0$; in experiments $\Delta t$ should be much smaller than the relaxation time of the system \cite{Brettschneider2011} and, in the presence of colored noise, should also meet the condition $\Delta t \gg \tau$. 

The symbols in Fig. \ref{fig2}(c) represent the experimental values of $S(y)$ for various $\sigma$ and $\tau$; they clearly show that, for the system described by Eq. (\ref{eq:SDEcircuit}), $S(y)$ is a constant that depends only on the intensity of the input noise $\sigma$, i.e. $S(y) = \frac{1}{2}\sigma^2$, and not on $\tau$. Fig. \ref{fig2}(d) shows the deterministic response $G(y)$ [solid line] and the experimental values of $D(y)$ [symbols]; the values of $D(y)$ lay on $G(y)$ independently of $\sigma$ and $\tau$. We note that the absence of dependence on $\tau$ for both $S(y)$ and $D(y)$ demonstrates that a white noise is a good model for the colored driving noise used in our experiments, i.e. with $\tau \leq 1.1\,\mathrm{\mu s}$.

\section{System with feedback}

\begin{figure}
\includegraphics[width=10cm]{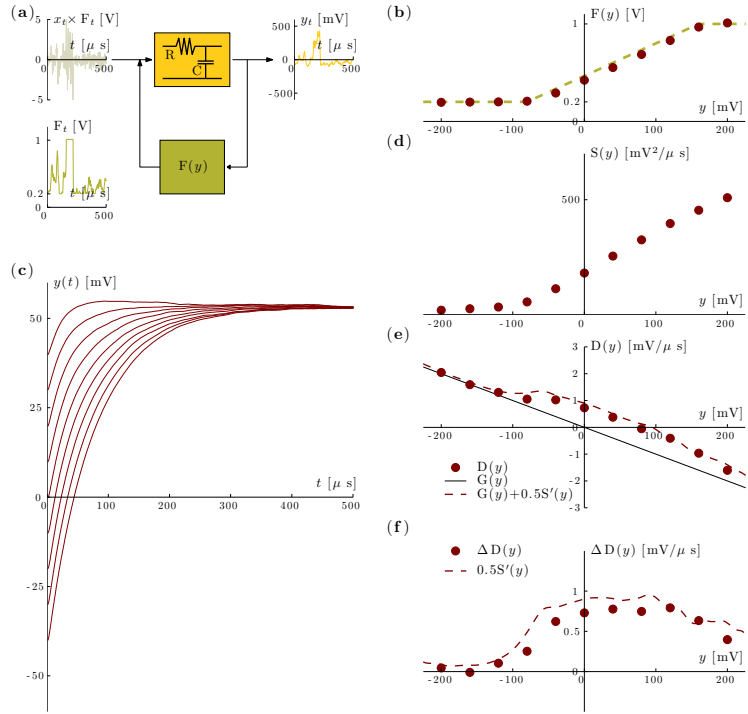}
\caption{{\bf Stochastic dynamical system with feedback.} (a) Schematic of a stochastical dynamical system with multiplicative feedback $F(y)$: the driving noise $x_t$ ($\tau = 1.1\,\mathrm{\mu s}$) is multiplied by a function of the system status $y_t$. (b) Nominal (dashed line) and experimentally measured (dots) feedback function used in our experiments. (c) Average of 1000 trajectories for various initial conditions; there is a clear shift of the equilibrium in comparison with the case without feedback [Fig. \ref{fig1}(c)]. (d) Diffusion $S(y)$ (dots) and (e) drift $D(y)$ (dots) of the system status. In (e), the solid line represents the harmonic restoring $G(y)$ and the dashed line $G(y)+0.5 S'(y)$. (f) Agreement between the noise-induced extra-drift $\Delta D(y)$ and $0.5 S'(y)$.\label{fig3}}
\end{figure}

Now we introduce a multiplicative feedback in the circuit as shown in Fig. \ref{fig3}(a). 
This is achieved by multiplying the input noise by $F(y)$. As shown in Fig. \ref{fig3}(b), $F(y)$ increases linearly between $-80\,\mathrm{mV}$ and $160\,\mathrm{mV}$ and saturates to $0.2\,\mathrm{V}$ ($1\,\mathrm{V}$) for $y <-80\,\mathrm{mV}$ ($y >160\,\mathrm{mV}$). The details of the circuit with multiplicative feedback are given in the methods section. The relevant SDE is:
\begin{equation}\label{eq:SDEmultcircuit}
dy_t = -k\,y_t\,dt + \sigma\,F(y_t)\,dW_t,
\end{equation}
which now requires an explicit specification of $\alpha$ in order to be well-defined. For the case of an electric circuit driven by a colored noise the Stratonovich convention holds, as is expected theoretically from the Wong-Zakai theorem \cite{Wong1969} and has been shown experimentally \cite{Smythe1983}. In fact, we see that the Stratonovich integral also describes the system in our case.

When $\tau = 1.1\,\mathrm{\mu s}$, differently from the case without feedback [Fig. \ref{fig2}(b)], the average trajectories in Fig. \ref{fig3}(c) do not converge to $y=0$, but to $y = 50\,\mathrm{mV}$. This shift of the equilibrium is a consequence of the non-uniformity of $S(y)$ [Fig. \ref{fig3}(d)] due to the presence of a multiplicative feedback. 

$D(y)$ [symbols in Fig. \ref{fig3}(e)] is also altered as a consequence of the multiplicative noise. In particular, $D(y)$ is now different from $G(y)$ [solid line in Fig. \ref{fig3}(e)]. The difference between the two is a noise-induced extra-drift
\begin{equation}\label{eq:spuriousdrift}
\Delta D(y) = D(y)-G(y),
\end{equation}
which is represented by the symbols in Fig. \ref{fig3}(f). We remark that $S(y)$ is independent from the interpretation of the underlying SDE \cite{Brettschneider2011}.

The relation between $\Delta D(y)$ and the variation of $S(y)$, i.e. $S'(y) = \frac{\partial S(y)}{\partial y}$, becomes evident considering the good agreement between $\Delta D(y)$ and $0.5 S'(y)$ [dashed line in Fig. \ref{fig3}(f)]. The prefactor $0.5$ corresponds to the $\alpha$ of the Stratonovich interpretation of the SDE (\ref{eq:SDEmultcircuit}), which permits us to make sense of the experimentally observed data.

We can therefore define
\begin{equation}\label{eq:alpha}
\alpha(y) = \frac{\Delta D(y)}{S'(y)},
\end{equation}
which in general may depend on the system under study \cite{vanKampen1981,Ao2007}.

\section{Dependence of $\alpha$ on $\tau/\delta$}

\begin{figure}
\includegraphics[width=10cm]{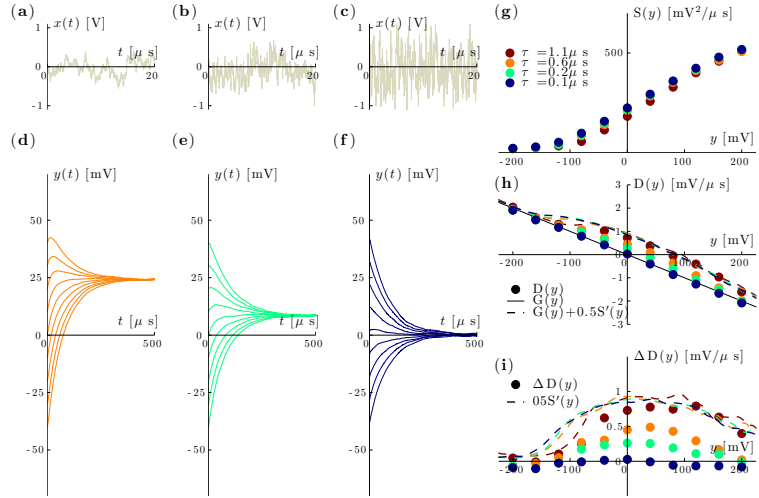}
\caption{{\bf Dependence on the noise correlation time $\tau$.} (a-c) Samples of input noises $x_t$ with $\tau = \mathrm{1.1,\, 0.6,\, 0.1\, s}$ for respectively (a), (b), (c). (d-f) Average of 1000 trajectories for various initial conditions with $\tau = \mathrm{1.1,\, 0.6,\, 0.1\, s}$ for respectively (d), (e), (f); there is a shift of the equilibrium towards $y=0$. (g) Dissusion $S(y)$. (h) Drift $D(y)$ (dots), harmonic restoring $G(y)$ (solid line) and $G(y)+0.5 S'(y)$ (dashed lines). (i) the ratio between the noise-induced extra-drift $\Delta D(y)$ and $0.5 S'(y)$ clearly varies as a function of $\tau$.
\label{fig4}}
\end{figure}

We now proceed to decrease $\tau$. Some samples of $x_t$ are shown in Figs. \ref{fig4}(a)-(c): the oscillations become faster and wider as $\tau$ decreases ($\tau = 0.6$, $0.2$ and $0.1\,\mathrm{\mu s}$ for Figs. \ref{fig4}(a), (b) and (c), respectively). We remark that the shorter the $\tau$, the more closely the conditions for the applicability of the Wong-Zakai theorem  \cite{Wong1969} are met. One might expect that the circuit equation will follow the Stratonovich equation even more closely and, thus, we shall expect no change with respect to the situation illustrated in Fig. \ref{fig3}. However, as we can see in Figs. \ref{fig4}(d)-(f), as $\tau$ decreases, the equilibrium position of the system moves back towards $y=0$.

Clearly, this behavior does not depend on the feedback; in fact, $F(y)$ is the same in all the cases, as evidenced by the fact that the experimental values of $S(y)$ do not vary significantly [symbols in Fig. \ref{fig4}(g)]. Instead, it depends on the fact that, as $\tau$ decrease, $D(y)$ [symbols in Fig. \ref{fig4}(h)] tends to $G(y)$ or, equivalently, $\Delta D(y)$ [symbols in Fig. \ref{fig4}(i)] tends to $0$. Using Eq. (\ref{eq:alpha}) and $S'(y)$ [dashed line in Fig. \ref{fig4}(i)], it is possible to calculate $\alpha$, which goes from $0.5$ to $0$ as $\tau$ decreases. Thus, the SDE (\ref{eq:SDEmultcircuit}) shifts from obeying the Stratonovich calculus ($\alpha = 0.5$ for $\tau = 1.1\,\mathrm{\mu s}$) to obeying the It\^o calculus ($\alpha = 0$ for $\tau = 0.1\,\mathrm{\mu s}$). As we have remarked  in the introduction, such a Stratonovich-to-It\^o transition can have dramatic effect on the long time dynamics of the system, for example, altering the system's equilibria a
 s shown in
  Figs. \ref{fig4}(d)-(f).

\begin{figure}
\includegraphics[width=10cm]{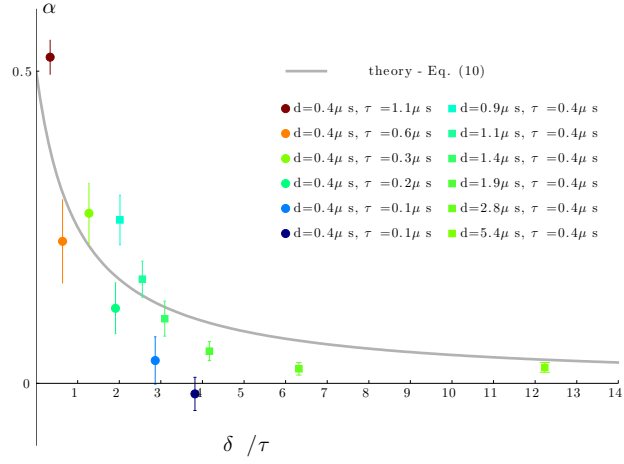}
\caption{{\bf Dependence of $\alpha$ on $\delta/\tau$.} $\alpha$ varies from 0.5 (Stratonovich integral) to 0 (It\^o integral) as $\delta/\tau$ increases. The solid line represents the results of the theory [Eq. (\ref{eq:mathe_alpha})]; the dots represent the values of $\alpha$ for fixed $\delta = 0.4\,\mathrm{\mu s}$ and varying $\tau$ [Fig. \ref{fig4}]; and the squares for fixed $\tau = 0.4\,\mathrm{\mu s}$ and varying $\delta$.
\label{fig5}}
\end{figure}

The reason for this Stratonovich-to-It\^o transition lies in the underlying dynamics of the system modeled by the SDE (\ref{eq:SDEmultcircuit}). For most real physical, chemical, biological and economic phenomena such microscopic dynamics are either too complex to be modeled or simply experimentally inaccessible. This justifies the need to resort to effective models, e.g. SDEs. For this work, however, we have chosen a model system, i.e. an electric circuit, that gives us complete access to the underlying dynamics. We are, therefore, able to track down the observed Stratonovich-to-It\^o transition to the fact that the feedback is not instantaneous, but entails a delay. We measured the feedback delay in the circuit in Fig. \ref{fig3}(a) to be $\delta = 0.4\,\mathrm{\mu s}$ (see methods section). The dots in Fig. \ref{fig5} represent $\alpha$ as a function of $\delta/\tau$. The transition occurs as $\tau$ becomes similar to $\delta$, i.e. $\delta/\tau \approx 1$. This can be qua
 litatively explained considering that, if $\delta = 0$, there is a correlation between the sign of $x$ and the time-derivative of $F(y)$, which is the underlying reason why the process converges to the Stratonovich solution \cite{Wong1969}; however, if $\delta \gg \tau$, this correlation disappears effectively randomizing the time-derivative of $F(y)$ with respect to the sign of $x$ and leading to a situation where the system loses its memory.

In order to gain a more precise mathematical understanding of this Stratonovich-to-It\^o transition, we consider the following family of delayed ODEs
\begin{equation}\label{eq:approx}
dy_t = -k\,y_t\,dt + \sigma\,F(y_{t-\delta})\,x_t^{\tau}dt,
\end{equation}
where $x_t^{\tau}$ is a sufficiently regular noise with correlation time $\tau$ and the feedback is delayed by $\delta$. Studying the limits where $\delta,\,\tau \rightarrow 0$ under the condition $\delta/\tau \equiv $ constant, we recover the SDE (\ref{eq:SDEmultcircuit}) with 
\begin{equation}\label{eq:mathe_alpha}
\alpha\left(\frac{\delta}{\tau}\right) = \frac{0.5}{1+\frac{\delta}{\tau}}.
\end{equation}
The details of this derivation are given in the methods section. Fig. \ref{fig5} shows the agreement between Eq. (\ref{eq:mathe_alpha}) [grey line] and the experimental data as a function of $\tau$ with fixed $\delta = 0.4\,\mathrm{\mu s}$ [dots]. 

In order to verify the dependence of $\alpha$ on the ratio $\delta/\tau$, we performed some additional experiments keeping $\tau = 0.4\,\mathrm{\mu s}$ fixed and varying $\delta$. For this purpose, we added a delay line in the feedback branch of the circuit so that we could adjust $\delta = 0.9$ to $5.4\,\mathrm{\mu s}$ (see methods section). The resulting values of $\alpha$ are plotted in Fig. \ref{fig5} as squares and are in good agreement with the theoretical prediction given by Eq. (\ref{eq:mathe_alpha}).

\section{Conclusion}

Our results show that the intrinsic ambiguity in the models of physical, biological and economical phenomena using SDE with multiplicative noise can have concrete consequences. In particular, even if an SDE with a specified convention is given, such convention can vary as a function of the hidden underlying dynamics of the system and therefore as a function of the position on the parameter space where the system is operated. Notably, our result that a Stratonovich-to-It\^o transition occurs if the delay in the feedback ($\delta$) is longer than the correlation time of the noise ($\tau$) has general applicability since instantaneous feedback and white noise are only mathematical approximations. The possibility of such a shift and of its dramatic consequences should be recognized and accounted for in many cases where SDEs with multiplicative noise are routinely employed to predict the behavior and evolution of complex physical, chemical, biological and economic phenomena.

\begin{appendix}
\section{RC Circuit}
The dynamical system employed in our experiments is an RC-electric circuit. A noisy signal $x_t$, which is generated by a function wave generator (Agilent 33250A) and pre-filtered by a low-pass filter to set the desired $\tau$, drives the RC series. The system's state $y_t$ is measured on the capacitor using a digital oscilloscope (Tektronix 5034B, $350\,\mathrm{MHz}$ bandwidth) at $10^6\,\mathrm{sample/s}$. For the circuit with feedback, a high-speed low-noise analog multiplier (AD835) is employed to multiply $x_t$ by the feedback signal (generated by amplifying $y_t$ and adding an offset) before applying it to the RC series. We measured the intrinsic delay of the circuit feedback branch (due to its finite bandwidth) applying a periodic deterministic signal and measuring the delay of the response. The additional delay line was realized by employing an analog variable delay amplifier (Ortec 427A).

\section{Derivation of Eq. (\ref{eq:mathe_alpha})}
We study the solution of Eq. (\ref{eq:approx}) taking the limit $\tau, \delta \rightarrow 0$ while keeping $\delta/\tau \equiv$ constant. In order to deal with a sufficiently regular process, we take $x_t^{\tau}$ as a harmonic process \cite{Schimansky-Geier1990}, i.e. the stationary solution of the SDE 
\begin{equation}\label{eq:harmonicnoise}
\left\{
\begin{array}{ccl}
dx_t^{\tau} & = & \frac{1}{\tau} z_t dt \\
dz & = & - \frac{\Gamma}{\tau} z_t dt - \frac{\Omega^2}{\tau} x_t^{\tau} dt + \frac{\sqrt{2 \gamma} \Omega^2}{\sqrt{\tau}} dW_t
\end{array}
\right.
\end{equation}
where $\Gamma$, $\Omega$ and $\gamma$ are constants, $W_t$ is a Wiener process, and $\tau$ is the correlation time for the Ornstein-Uhlenbeck process obtained taking the limit $\Gamma,\, \Omega ^2 \rightarrow \infty$ while keeping $\frac{\Gamma}{\Omega ^2} = 1$. As $\tau \rightarrow 0$, the rescaled solution of Eq. (\ref{eq:harmonicnoise}) $\frac{x_t}{\sqrt{\tau}} $ converges to a white noise.\\
In Eq. (\ref{eq:approx}), we make the time substitution $u = t-\delta$ and then write the equation in terms of the Wiener process $V_u$ defined as $V_u = W_{u + \delta} - W_{\delta}$.  Next, we expand about $u$ to first order and rewrite the resulting equation as a first order system in $y$, $v$, $x$, and $z$, where $v = \sqrt{\frac{\delta}{\tau}}\sqrt{\tau}\frac{dy}{du}$.  We then consider the backward Kolmogorov equation associated with the resulting SDE, which gives the equation for the transition density $\rho(u, y, v, x, z, y', v', x', z', u')$. We can expand $\rho$ in powers of the parameter $\sqrt{\tau}$, i.e. $\rho = \rho_0 + \sqrt{\tau} \rho_1 + \tau \rho_2 + ...$. We use the standard homogenization method \cite{Pavliotis} to derive the backward Kolmogorov equation for $\rho_0$ \cite{Risken}, i.e. the equation for the the limiting transition density $\rho_0$ as $\tau, \delta \rightarrow 0$ with $\delta/\tau \equiv$ constant. Finally, we take the limit $\Gamma, \Omega
  ^
 2 \rightarrow \infty$ while keeping the ratio $\frac{\Gamma}{\Omega ^2} = 1$. The resulting backward Kolgomorov equation is 
\begin{equation}\label{eq:Kolgomorov}
\frac{\partial \rho _0}{\partial u} = 
\left[ 
- k y 
+ 
\frac{0.5}{1+\frac{\delta}{\tau}}
\sigma^2 F(y) \frac{dF(y)}{dy}
\right] 
\frac{\partial \rho _0}{\partial y}
+ 
\frac{\sigma^2 F^2(y)}{2}
\frac{\partial ^2 \rho _0}{\partial y ^2},
\end{equation}
and the associated (It\^o) SDE is 
\begin{equation}\label{eq:SDEKolgomorov}
dy_t =  - k y_t dt 
+ 
\frac{0.5}{1+\frac{\delta}{\tau}}
\sigma^2 F(y) \frac{dF(y)}{dy}
dt
+ 
\sigma F(y) dW_t.
\end{equation}
The equation for $\alpha$ [Eq. (\ref{eq:mathe_alpha})] follows straightforwardly by comparison of Eq. (\ref{eq:SDEKolgomorov}) and Eq. (\ref{eq:SDEmult}).

\begin{acknowledgments}
The authors would like to thank Clemens Bechinger, Laurent Helden, Ao Ping, Riccardo Mannella and Antonio Sasso for inspiring discussions, Sergio Ciliberto and Antonio Coniglio for critical reading of the manuscript, and Alfonso Boiano for help in the realization of the circuit.
\end{acknowledgments}

\bibliographystyle{unsrt} 
\bibliography{biblio}

\end{appendix}



\end{document}